\documentclass[11pt,twoside]{article}
\usepackage{./asp2014}

\aspSuppressVolSlug
\resetcounters

\begin{document}

\title{Science with the ngVLA: Planetary Nebulae}
\author{Joel H. Kastner,$^1$, Albert Zijlstra,$^2$, Bruce Balick$^3$, and Raghvendra Sahai$^4$
\affil{$^1$Rochester Institute of Technology, Rochester, NY, USA; \email{jhk@cis.rit.edu}}
\affil{$^2$University of Manchester, Manchester, UK; \email{a.zijlstra@manchester.ac.uk}}
\affil{$^3$University of Washington, Seattle, WA, USA; \email{balick@uw.edu}}
\affil{$^4$JPL, Pasadena, CA, USA; \email{raghvendra.sahai@jpl.nasa.gov}}}

\paperauthor{Joel H. Kastner}{jhk@cis.rit.edu}{0000-0002-3138-8250}{Rochester Institute of Technology}{Center for Imaging Science}{Rochester}{NY}{14623}{USA}
\paperauthor{Raghvendra Sahai}{raghvendra.sahai@jpl.nasa.gov}{ORCID_Or_Blank}{JPL}{Author2 Department}{Pasadena}{CA}{XXXXX}{USA}
\paperauthor{Albert Ziljstra}{a.zijlstra@manchester.ac.uk}{0000-0002-3171-5469}{University of Manchester}{Author3 Department}{Manchester}{State/Province}{Postal Code}{UK}
\paperauthor{Bruce Balick}{balick@uw.edu}{ORCID_Or_Blank}{University of Washington}{Dept.\ of Astronomy}{Seattle}{WA}{XXXXX}{USA}

\begin{abstract}
Planetary nebulae (PNe) represent the near endpoints of evolution for stars of initial mass $\sim$1--8 $M_\odot$, wherein  the envelope of an asymptotic giant branch (AGB) star becomes photodissociated and ionized by high-energy radiation from a newly emerging white dwarf that was the progenitor star's core. It is increasingly evident that most PNe are descended from binary systems. PNe hence provide unique insight into a diverse range of astrophysical phenomena, including the influence of companion stars on the late stages of stellar evolution; stellar wind interactions and shocks; the physics and chemistry of photoionized plasmas and photon-dominated regions (PDRs); and enrichment of the ISM in the products of intermediate-mass stellar nucleosynthesis. We describe specific examples of the potential impact of the ngVLA in each of these areas.
\end{abstract}

\section{Introduction}

Planetary nebulae (PNe) serve as textbook examples of astrophysical plasmas and shock processes and provide essential tests of theories of stellar evolution and the origin and enrichment of the heavy elements in the universe \citep[e.g.,][]{Kwok2000}.  Classically, a PN results from the ejection and ionization of the envelope of an asymptotic giant branch (AGB) star --- the descendant of a progenitor of initial mass $\sim$1--8 $M_\odot$ \citep[see chapter by][in this volume]{MatthewsSciBook} --- wherein the source of ionizing (UV) photons is a newly emerging white dwarf that was the progenitor star's core. PNe also exhibit a dazzling variety of aspherical morphologies: elliptical; bipolar or multipolar; highly point-symmetric; chaotic and clumpy \citep{Sahai2011}. The physical mechanisms responsible for this PN morphological menagerie have been the subject of intense interest and hot debate among PN researchers for well over two decades
\citep[e.g.,][and references therein]{BalickFrank02, Akashi2018}. The PN shaping problem is multifaceted, with connections to (and implications for) a wide variety of astrophysical systems. Areas of particular importance are binary star astrophysics; stellar wind interactions and shocks; the physics and chemistry of photoionized plasmas and photon-dominated regions (PDRs); and enrichment of the ISM in the products of intermediate-mass stellar nucleosynthesis.  The study of PNe with the ngVLA will have major impact in all of these areas. Here, we offer a few prime examples.


\section{Recombination lines: nebular dynamics and abundances of light elements \label{sec:recomb}}

The radio spectra of planetary nebulae include a large number of bound-bound recombination lines. The brightest lines are those of the H$n\alpha$ series. The helium lines are a factor of 10 fainter, and the carbon lines fainter still. Beyond this, the lines from the different elements merge. These radio recombination lines provide a powerful diagnostic of PN internal regions that is not affected by extinction, and is much less temperature sensitive than optical forbidden lines. Radio lines provide a 2-D map of PN internal dynamics (from the line profiles), the local electron temperatures (from the line to continuum ratios), and the helium abundance (from the helium to hydrogen line ratios).  The power of these methods was shown by \citet{Roelfsema1991} for the planetary nebula NGC 7027.

For an HST-like resolution of 0.1$''$, the ngVLA has a line sensitivity (1-sigma) in 1 hour of $\sim$100 microJy beam$^{-1}$, assuming a linewidth of 10 km s$^{-1}$ that is appropriate for PNe like NGC 7027 (Fig.~\ref{fig:N7027}). The line to continuum ratio is typically 10 per cent, increasing as $\nu^{1.1}$. Young planetary nebulae have brightness temperatures of $T_{\rm b}=10^3$\,K at 5\,GHz, with a continuum flux of around 5 mJy beam$^{-1}$, and $\sim$50\% half have $T_{\rm b} \stackrel{>}{\sim} 10^2$\,K \citep {vandeSteene1995}. Hence ngVLA mapping of radio recombination lines would be feasible even at this extreme resolution. In fainter regions, lower angular resolution can be used, and multiple recombination lines within the observed frequency band can be summed to increase S/N. Higher angular resolution would not be useful; for thermal lines, most collecting area should be within the inner 100 km.

The continuum flux declines only as $\nu^{-0.1}$ towards higher frequencies, while the recombination lines increase in strength and the ngVLA sensitivity at constant resolution is fairly constant. The recombination lines can thus be mapped over the full frequency range available to the ngVLA, which spans transitions from H40$\alpha$ to well over H120$\alpha$, with reduced sensitivity above 50 GHz due to the higher angular resolution. These observations would provide unique detail concerning the 2-D velocity fields of PNe, while the velocity-depth profile can be traced by observing at frequencies where the nebula becomes optically thick. Deviations from LTE can be measured; at low $n$, the  H$n\alpha$ lines are sensitive to laser amplification within circumstellar disks at the cores of PNe that are of particular interest for shaping studies \citep[][see \S \ref{sec:binaries}]{Aleman2018}.

\section{Precise expansion velocities and distances from nebular proper motions}

The fundamental parameters of PNe --- such as emission line and continuum luminosities and linear dimensions --- are generally poorly determined, thanks to the notorious difficulty in ascertaining PN distances. Even in the era of {\it Gaia}, trigonometric parallax distances to all but the nearest and brightest PNe will likely remain uncertain, given the challenges inherent in obtaining precise astrometry for faint and/or heavily optically obscured central stars or diffuse PN emission. The ngVLA will enable the broad application of the expansion parallax distance determination technique, which  was pioneered via VLA and HST studies of relatively nearby PNe and compact H {\sc ii} regions \citep[e.g.,][]{Masson1989, Acord1998,Palen2002,Guzman2006}, to the hundreds of Galactic PNe recently discovered via, e.g., large-area H$\alpha$ surveys \citep[][]{Frew2013,Sabin2014}. With angular resolution a factor $\sim$10 times better than that of the seminal \citet{Masson1989} study of NGC 7027 ($D=890$ pc) over the same (5--15 GHz) frequency range, the ngVLA can provide expansion parallax distances accurate to $\sim$10\% for PNe out to $\sim$10 kpc, with the potential for much higher precision distance determinations for objects in the solar neighborhood. 

Conversely, the precise measurement of proper motions of nebular gas for the significant subset of nearby PNe that have bright central stars --- and, hence, will have well-determined {\it Gaia} distances --- offers an entirely new approach to modeling PN bulk gas expansion and the short-timescale evolution of ionization fronts. 
A planetary nebula is shaped as faster and highly directed stellar outflows from the PN core and/or a binary companion (see \S~\ref{sec:binaries}) supersonically overtake the much more massive, isotropic, slow winds deposited over thousands of years as the star ascended the AGB. The geometry and momentum flux of the fast, collimated wind evolve as the CSPN traverses the H-R diagram from the tip of the AGB ($T_{eff} \sim 3500$ K) to the white dwarf turnoff ($\sim$200 kK) at constant luminosity ($L \sim 10^{3.6} L_\odot$). The fast-slow wind interactions produce shock interfaces of varying shapes that expand at Mach 3--10 (30--100 km s$^{-1}$) into the ambient AGB gas upstream. While the shocks themselves are only detectable, and only under specific conditions, via X-rays \citep{Kastner2012}, the sharp (spatially unresolved) outer edges of a bright rim of wind-displaced and compressed slow gas mark their locations. The movements of this rim can be traced via observations with geometrically stable imaging instruments with $\le$100 mas spatial resolution \citep[e.g.,][]{Schoenberner2018} --- a requirement readily satisfied by the ngVLA. Indeed, the long-baseline synthesis imaging enabled by the ngVLA will be the tool of choice for tracking the progress of shock interfaces after HST has reached mission end. 

The changes of rim shape measured via multi-epoch ngVLA imaging, coupled with ever-improving hydro simulations of the rim formation process and new model atmosphere calculations of the evolving CSPN, will allow us to retrace the histories of the stellar winds in a strategically selected sample of bright, nearby targets. For these objects, we can obtain both the tangential expansion speeds and radial velocities (via radio or optical recombination line profile modeling; \S~\ref{sec:recomb}). This will effectively eliminate a major source of uncertainty in ascertaining the three-dimensional structures of PNe, leading to radical advances in our understanding of PN shaping. 
Such ngVLA studies can be carried out at wavelengths 2 to 4 cm in all but the worst of local weather.   In contrast, adaptive optics (AO) and interferometric near-IR imaging aimed at the same science is affected by a spatially complex mixture of scattering small dust and emitting cool dust that is difficult to disentangle, and the PSFs of optical/near-IR AO instruments tend to be too sensitive to atmospheric conditions for purposes of precise measurements of proper motions of complicated nebular patterns over few-year temporal baselines.  

\section{Nebular structures and substructures: diagnosing close binary interactions \label{sec:binaries}}

As noted by \citet{MatthewsSciBook}, there is now broad agreement that the transformation from a quasi-spherical wind during the progenitor star's AGB phase to nonspherical or highly collimated outflows during post-AGB, pre-PN, and PN phases can be traced to the influence of a companion to the mass-losing central star \citep[][and references therein]{Frank2018,Akashi2018}. This consensus was forged on the basis of two decades of (mostly HST) high-resolution imaging which, when placed in the context of detailed simulations of PN morphological evolution, has shown convincingly that PNe are initially shaped, during late AGB and post-AGB (pre-PN) stages, from the inside out by fast, collimated outflows \citep{SahaiTrauger1998,HuaEsp2012}. Such collimated outflows are a natural consequence of interacting binary systems, whether via common envelope (CE) evolution or jets associated with a companion star's accretion disk. Under either scenario, the highly directed outflows power their way through through the far more massive but more spherically symmetric remains of slow winds previously produced from the parent AGB star.  After a few centuries these jets produce the observed pairs of lobes that characterize bipolar/multipolar pre-PNe and PNe. The sculpting action of jets may also be responsible for weaker departures from sphericity apparent in the vast majority of (elliptical) PNe \citep{Soker1990,HuaEsp2012}. 

This realization is forcing a paradigm shift in the goals and strategies of observational programs that can be undertaken over the upcoming decade with radio and IR telescopes around the world and in orbit.  The effects of the formation of accretion disks and common envelope evolution are arguably more readily studied in pre-PNe and PNe than in SNe Ia, novae, and cataclysmic binaries; compared to these exciting but highly transient sources, the binary system outflows presented by PNe and their progenitors are closer, far more numerous, and slower to unfold.  Spectroscopic instruments, such as the ngVLA, that combine very high spatial precision and larger collecting area with geometric stability can exploit the resulting opportunities to trace wind-sculpted density structures over a wide range of size scales, as well as monitor small-scale structural evolution over long temporal baselines, for signficant samples of pre-PNe and PNe.

\articlefigure{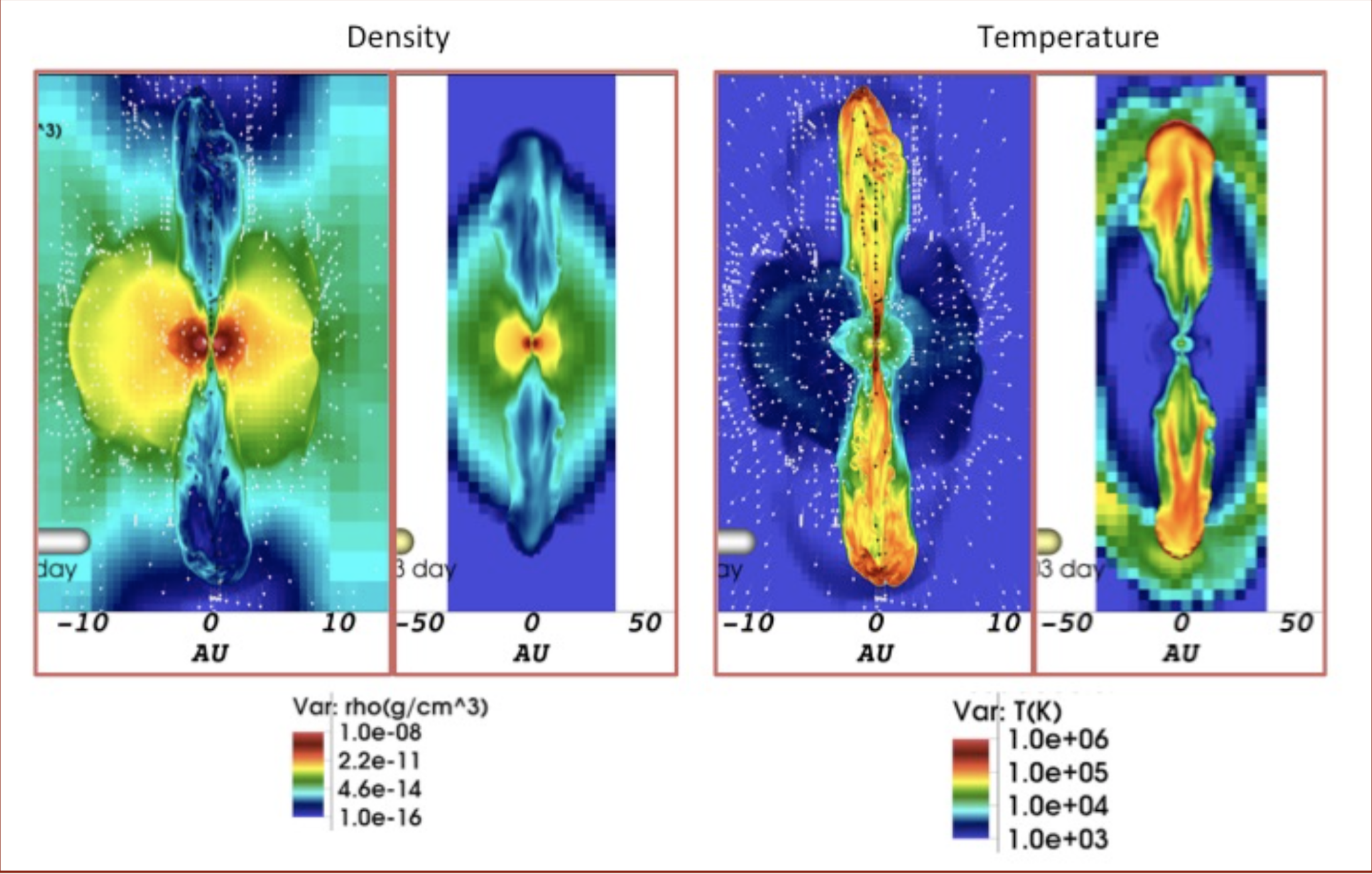}{fig:binaryModel}{Results of 3-D adaptive mesh refinement simulations of a high-momentum fast wind being driven into the ejecta resulting from the onset of a binary star common envelope event. The left and right panels illustrate the distributions of density and temperature, respectively, at $t = 780$ days (left subpanels) and $t = 1600$ days (right subpanels) after initiation of the fast wind. Note the change of scale between the two times as the nebular lobes expand. Figure from \citet{Frank2018}.
}

The sensitivity and resolving power of the ngVLA will finally provide the data needed to inform and test AGB and post-AGB ejecta (hence PN) shaping models. Most such models invoke Roche lobe overflow resulting in companion accretion disk and jet formation \citep{Akashi2018} or wind interactions resulting from the sudden onset of the CE phase of a close (AGB + MS star) binary system \citep{Frank2018}. A simulation of the collimated outflows emerging within a few years of the onset of CE is presented in Fig.~\ref{fig:binaryModel}. The characteristic size scales of the outflow collimator (i.e., the CE ejecta) and the wind interaction region are $\sim$10--50 AU, requiring resolution of order 10 mas given typical post-AGB and PNe distances (a few kpc). The ngVLA will have unparalleled capability to probe such size scales via 10--100 GHz imaging. The collimating structures (i.e., disks and tori) can be traced in thermal continuum emission (i.e., Bremsstrahlung and/or emission from large dust grains) as well as CO isotopologues and dense gas tracers (CS, HCN, HCO$^+$). Subarcsecond low-$T_{\rm ex}$ molecular line imaging with the ngVLA will complement ALMA studies in distinguishing kinematically between Keplerian disks, expanding tori, and outflow structures \citep[e.g.,][]{Bujarrabal2018}. The potent combination of ngVLA and ALMA will hence represent the tool of choice for studies of AGB or post-AGB stars that are the suspected sites of recent CE events, or the cores of young PNe whose complex large-scale morphologies, evidence for jets, etc., makes them the suspected hosts of interacting binaries.
 
\section{Central stars: binaries and winds}

Given the importance of central binaries in PN shaping, concerted observational effort has been devoted to searches for companions to PN central stars in recent years (via, e.g., central star photometric monitoring), resulted in numerous detections \citep{Hillwig2018}. However, such optical survey work remains limited to relatively bright and nearby central stars. Meanwhile, roughly 50\% of nearby PNe show an X-ray source at the cental star, and most of these point-like X-ray sources appear too hard to be due to shock processes \citep{Kastner2012}.  The most likely explanation is that a low-mass companion star has accreted material from the primary during AGB mass loss and subsequent PN ejection, and has been spun up; the resulting fast rotation reactivates the secondary's internal magnetic dynamo, leading to coronal emission. This model predicts coronal radio emission as well, if such stars follow the Guedel-Benz relation between X-ray luminosity and radio luminosity density \citep[e.g.,][]{Forbrich2011}. Indeed, pilot e-Merlin observations have detected radio continuum emission from LoTr 5 --- a PN (at $D=500$ pc) known to harbor a central binary that is a luminous, point-like X-ray source --- at a level consistent with the flux predicted from the X-rays. This demonstrates the potential of the ngVLA, which can be used to efficiently survey all nearby PNe to search for central star companions. 

The mass-losing central stars of planetary nebulae themselves have not yet been detected in the radio. They are expected to show weak emission from stellar winds, in addition to chromospheric emission from spun-up companions. Stellar winds are the largest uncertainty in the evolution  of PNe: they remove some of the remaining hydrogen envelope and thereby speed up the evolution towards higher temperatures.  These winds will finally be measurable with the ngVLA. Using the equations in \citet{WrightBarlow1975} for typical central star wind parameters
(a mass loss rate of $10^{-8}\,\rm M_\odot\,yr^{-1}$ and wind speed of $v_{\rm w}
=10^3\,\rm km\,s^{-1}$), and assuming a distance of 1\,kpc, the predicted flux at 5 GHz is 0.2 $\mu$Jy, increasing as $\nu^{0.7}$. Such wind emission becomes detectable at 40\,GHz; PNe at  smaller distances would be detectable at lower frequencies or mass loss rates. The aforementioned survey of PN central stars hence would not only have the potential to directly detect low-mass binary companions, but would resolve such companions from primaries that are themselves wind radio emission sources at separations as small as $\sim$100 AU. Such capabilities will be far superior to those of other (e.g., Chandra X-ray) central star wind and companion detection techniques. 

\section{Understanding physical and chemical processes in irradiated molecular gas}

Though best known as $\sim10^4$ K optical emission line sources, a subset of PNe retain significant masses of cold ($<100$ K), dense ($\sim10^4-10^6$ cm$^{-3}$) molecule-rich AGB ejecta  \citep[e.g.,][]{Huggins2005}.
This molecular gas is irradiated from within by UV from hot (30--200 kK) PN central stars.  A majority of PNe also display point-like X-ray emission from their central stars (or companions) and/or diffuse X-ray emission from rarefied, shock-heated ($T > 10^6$ K) plasmas \citep{Kastner2012}.  Such extremes of physical conditions within individual, readily resolvable objects with well-defined molecular gas irradiation geometries make PNe particularly fertile ground for the study of radiation-driven molecular gas heating and chemistry. The lessons learned can be applied to PDR environments ranging in scale from individual protoplanetary disks orbiting T Tauri stars to H {\sc ii} regions associated with massive young star clusters. 

\articlefigure{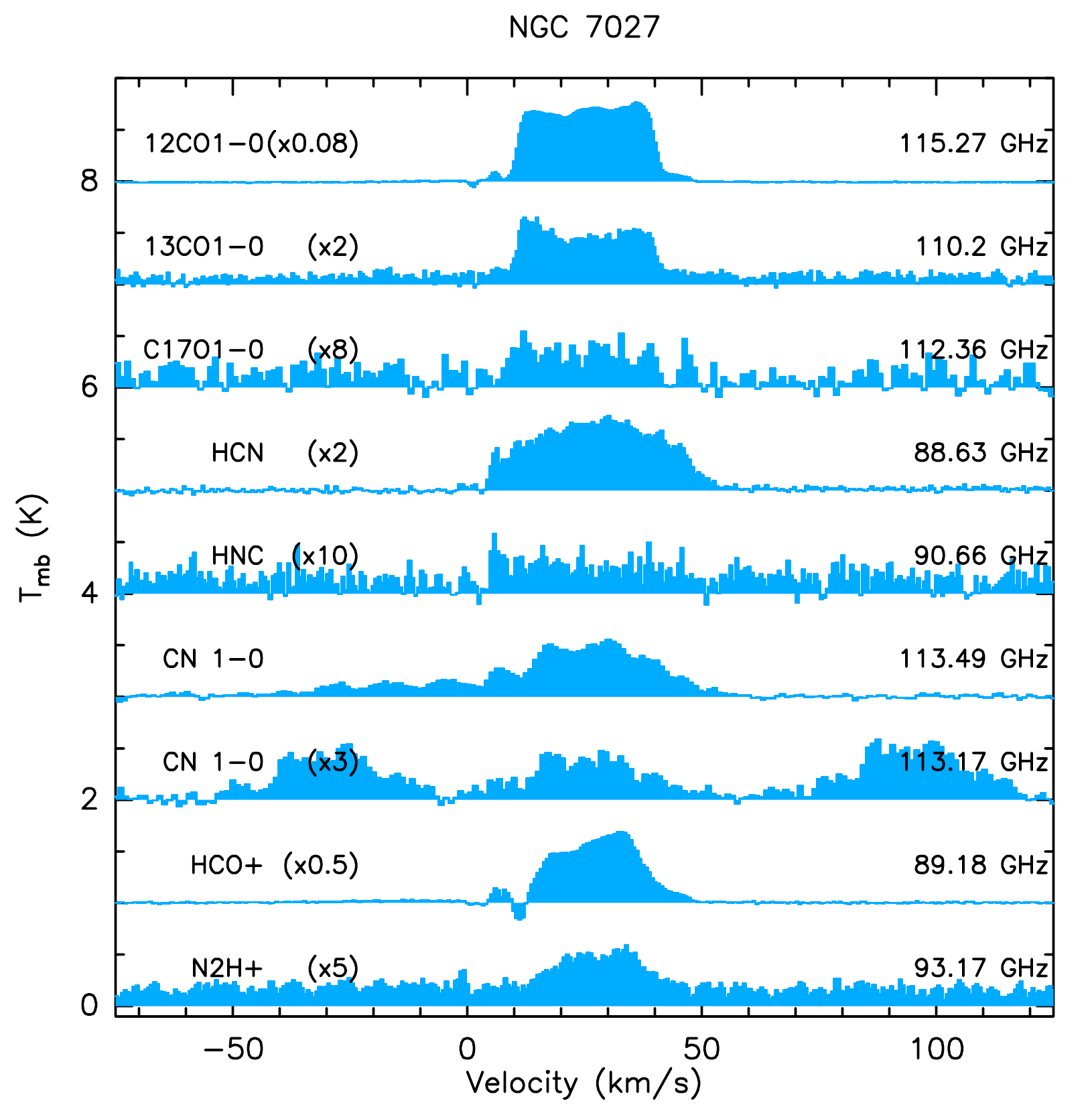}{fig:N7027}{Selected spectra from a recent single-dish (IRAM 30 m telescope) molecular line survey of planetary nebula ``poster child'' NGC 7027 \citep[adapted from][]{Bublitz2018}. All of these transitions will be accessible to the ngVLA's Band 6 (70--116 GHz).}

In PNe, as in any PDR environment, 
FUV and EUV photons are readily absorbed by neutral H and dust (respectively) in molecular surface layers, strongly influencing the heating of the gas in these layers. X-rays penetrate to larger depths, where they can ionize H$_2$ molecules as well as abundant atoms, such as He, N, and O, that have relatively high ionization potentials or tightly bound inner-shell electrons; the presence of highly reactive and/or abundant ions (e.g., H$_3^+$, He$^+$, N$^+$) can then lead to rich molecular chemistries via a miriad of reaction pathways.  The molecules HCO$^+$, N$_2$H$^+$, HCN, HNC, CN, and C$_2$H, along with radio tracers of H$_2$ density (i.e., CO and its isotopologues), represent particularly potent diagnostics of the effects and relative penetration depths of FUV/EUV and X-ray photons. 
The potential of PNe in this regard is apparent from the single-dish and interferometric molecular line surveys that have been carried out thus far for a relatively small number of objects \citep[e.g.,][see Fig.~\ref{fig:N7027}]{Bachiller1997,SchmidtZiurys2017a,SchmidtZiurys2017b,Bublitz2018}. 

The leap in frequency coverage of the ngVLA will unlock this potential: all of the transitions in Fig.~\ref{fig:N7027} will be accessible to the ngVLA's Band 6 (70--116 GHz) at sensitivities and spatial resolution exceeding that of ALMA by factors of $\sim$5 and $\sim$10, respectively. Thus, the ngVLA will greatly expand the number of PNe that can be surveyed in the aforementioned species; molecular line surveys of objects analogous to the well-studied, molecule-rich PNe in the solar neighborhood (e.g., the Helix, Ring, Dumbbell, NGC 7027, etc.) can be undertaken for objects as distant as the Galactic Bulge. Furthermore, the ngVLA will facilitate spatially resolved studies of the same species in many dozens of PNe for which only single-dish molecular line data exist and/or only CO or (near-IR) H$_2$ emission has been detected thus far \citep{Kastner1996,Huggins2005}. We can thereby measure the spatial gradients of temperature- and ionization-sensitive molecular line ratios (e.g., HNC/HCN, CN/HCN, HCO$^+$/CO) as functions of distance from UV- and X-ray-luminous PN central stars so as to constrain molecular gas ionization and photodissociation rates and inform models of UV- and X-ray-driven molecular chemistry in H {\sc ii} region PDRs and protoplanetary disks \citep[e.g.,][]{Graninger2014,Cleeves2015}. 

Finally, the exquisite sensitivity of the ngVLA will be especially conducive to observations of isotopologues of CN and HCN in AGB stars, pre-PNe, and PNe aimed at measurements of the isotopic ratios $^{12}$C/$^{13}$C and $^{14}$N/$^{15}$N \citep{Lee2013}. Ascertaining the range of these ratios in major sources of ISM C and N enrichment, like PNe, is key to addressing longstanding questions concerning Galactic chemical evolution and the origins of pre-solar grains \citep[e.g.,][and references therein]{HilyBlant2017,Ziurys2017}.






\begin{thebibliography}{}
\bibitem[Acord et al.(1998)]{Acord1998} Acord, J.~M., Churchwell, E., \& Wood, D.~O.~S.\ 1998, \apjl, 495, L107 
\bibitem[Akashi et al.(2018)]{Akashi2018} Akashi, M., Bear, E., \& Soker, N.\ 2018, \mnras, 475, 4794 
\bibitem[Aleman et al.(2018)]{Aleman2018} Aleman, I., Exter, K., Ueta, T., et al.\ 2018, \mnras, 477, 4499 
\bibitem[Bachiller et al.(1997)]{Bachiller1997} Bachiller, R., Forveille, T., Huggins, P.~J., \& Cox, P.\ 1997, \aap, 324, 1123 
\bibitem[Bublitz et al.\ (2018)]{Bublitz2018} Bublitz, J., Kastner, J.H., Santander-Garcia, M., et al. 2018, in preparation
\bibitem[Bujarrabal et al.(2018)]{Bujarrabal2018} Bujarrabal, V., Alcolea, J., Miko{\l}ajewska, J., Castro-Carrizo, A., \& Ramstedt, S.\ 2018, \aap, 616, L3 
\bibitem[Balick \& Frank (2002)]{BalickFrank02} Balick, B., \& Frank, A., 2002, ARAA, 40, 439
\bibitem[Cleeves et al.(2015)]{Cleeves2015} Cleeves, L.~I., Bergin, E.~A., Qi, C., Adams, F.~C., \& {\"O}berg, K.~I.\ 2015, \apj, 799, 204 
\bibitem[Frank et al.(2018)]{Frank2018} Frank, A., Chen, Z., Reichardt, T., et al.\ 2018, in Proceedings of ``Asymmetrical Planetary Nebular VII'' (special issue of {\it Galaxies}), in press; arXiv:1807.05925 
\bibitem[Frew et al.(2013)]{Frew2013} Frew, D.~J., Boji{\v c}i{\'c}, I.~S., \& Parker, Q.~A.\ 2013, \mnras, 431, 2 
\bibitem[Forbrich et al.(2011)]{Forbrich2011} Forbrich, J., Wolk, S.~J., G{\"u}del, M., et al.\ 2011, 16th Cambridge Workshop on Cool Stars, Stellar Systems, and the Sun, 448, 455 
\bibitem[Graninger et al.(2014)]{Graninger2014} Graninger, D.~M., Herbst, E., {\"O}berg, K.~I., \& Vasyunin, A.~I.\ 2014, \apj, 787, 74 
\bibitem[Guzm{\'a}n et al.(2006)]{Guzman2006} Guzm{\'a}n, L., G{\'o}mez, Y., \& Rodr{\'{\i}}guez, L.~F.\ 2006, RMxAA, 42, 127 
\bibitem[Hillwig(2018)]{Hillwig2018} Hillwig, T.\ 2018, Galaxies, 6, 85 
\bibitem[Hily-Blant et al.(2017)]{HilyBlant2017} Hily-Blant, P., Magalhaes, V., Kastner, J., et al.\ 2017, \aap, 603, L6 
\bibitem[Huarte-Espinosa et al.(2012)]{HuaEsp2012} Huarte-Espinosa, M., Frank, A., Balick, B., et al.\ 2012, \mnras, 424, 2055 
\bibitem[Huggins et al.(2005)]{Huggins2005} Huggins, P.~J., Bachiller, R., Planesas, P., Forveille, T., \& Cox, P.\ 2005, \apjs, 160, 272 
\bibitem[Kastner et al.(1996)]{Kastner1996} Kastner, J.~H., Weintraub, D.~A., Gatley, I., Merrill, K.~M., \& Probst, R.~G.\ 1996, \apj, 462, 777 
\bibitem[Kastner et al.(2012)]{Kastner2012} Kastner, J.~H., Montez, R., Jr., Balick, B., et al.\ 2012, \aj, 144, 58 
\bibitem[Kwok(2000)]{Kwok2000} Kwok, S.\ 2000, {\it The Origin and Evolution of Planetary Nebulae}. New York: Cambridge University Press~(Cambridge Astrophysics Series; 33)  
\bibitem[Lee et al.(2013)]{Lee2013} Lee, C.-F., Yang, C.-H., Sahai, R., \& S{\'a}nchez Contreras, C.\ 2013, \apj, 770, 153 
\bibitem[Masson(1989)]{Masson1989} Masson, C.~R.\ 1989, \apj, 336, 294 
\bibitem[Matthews \& Claussen (2018)]{MatthewsSciBook} Matthews, L. \& Claussen, C. 2018, ngVLA Science Book
\bibitem[Palen et al.(2002)]{Palen2002} Palen, S., Balick, B., Hajian, A.~R., et al.\ 2002, \aj, 123, 2666 
\bibitem[Roelfsema et al.(1991)]{Roelfsema1991} Roelfsema, P.~R., Goss, W.~M., Pottasch, S.~R., \& Zijlstra, A.\ 1991, \aap, 251, 611 
\bibitem[Sabin et al.(2014)]{Sabin2014} Sabin, L., Parker, Q.~A., Corradi, R.~L.~M., et al.\ 2014, \mnras, 443, 3388 
\bibitem[Sahai \& Trauger(1998)]{SahaiTrauger1998} Sahai, R., \& Trauger, J.~T.\ 1998, \aj, 116, 1357 
\bibitem[Sahai et al.\ (2011)]{Sahai2011} Sahai, R., Morris, M., \& Villar 2011, AJ, 141, 134 
\bibitem[Sch\"onberner et al.(2018)]{Schoenberner2018} Sch\"onberner, D, Balick, B., \& Jacob, R. 2018, A\&A 609, A126.
\bibitem[Schmidt \& Ziurys(2017a)]{SchmidtZiurys2017a} Schmidt, D.~R., \& Ziurys, L.~M.\ 2017a, \apj, 835, 79 
\bibitem[Schmidt \& Ziurys(2017b)]{SchmidtZiurys2017b} Schmidt, D.~R., \& Ziurys, L.~M.\ 2017b, \apj, 850, 123 
\bibitem[Soker(1990)]{Soker1990} Soker, N.\ 1990, \aj, 99, 1869 
\bibitem[van de Steene \& Zijlstra(1995)]{vandeSteene1995} van de Steene, G.~C., \& Zijlstra, A.~A.\ 1995, \aap, 293, 541 
\bibitem[Wright \& Barlow(1975)]{WrightBarlow1975} Wright, A.~E., \& Barlow, M.~J.\ 1975, \mnras, 170, 41 
\bibitem[Ziurys et al.(2017)]{Ziurys2017} Ziurys, L.~M., Woolf, N.~J., Schmidt, D.~R., Zack, L.~N., \& Zega, T.\ 2017, 80th Annual Meeting of the Meteoritical Society, 1987, 6408 

\end{thebibliography}


\end{document}